\documentstyle{article}

\newlength{\dinwidth}
\newlength{\dinmargin}
\makeatletter

\new@fontshape{msb}{m}{n}{%
   <5>msbm5%
   <6>msbm6%
   <7>msbm7%
   <8>msbm8%
   <9>msbm9%
   <10>msbm10%
   <11>msbm10%
   <12>msbm10 at10.95pt%
   <14>msbm10 at14.4pt%
   <17>msbm10 at17.28pt%
   <20>msbm10 at20.74pt%
   <25>msbm10 at24.88pt}{}
\extra@def{msb}{}{}
\newmathalphabet*\Bbb{msb}{m}{n}

\new@fontshape{euf}{m}{n}{%
   <5>eufm5%
   <6>eufm6%
   <7>eufm7%
   <8>eufm8%
   <9>eufm9%
   <10>eufm10%
   <11>eufm10 at10.95pt%
   <12>eufm10 at12pt%
   <14>eufm14%
   <17>eufm14 at17.28pt%
   <20>eufm14 at20.74pt%
   <25>eufm14 at24.88pt}{}
\extra@def{euf}{\hyphenchar#1\m@ne
       \@tempdima\fontdimen2#1%
       \fontdimen3#1.4\@tempdima
       \fontdimen4#1.3\@tempdima}{}%
\newmathalphabet*\got{euf}{m}{n}

\newif\if@fewtab\@fewtabtrue

\def\cases#1{\left\{\,\vcenter{\normalbaselines\m@th
    \ialign{$\displaystyle{##}\hfil$&\quad##\hfil\crcr#1\crcr}}\right.}

\makeatother

\def\R{{\Bbb R}}

\def\Z{{\Bbb Z}}
\def\N{{\Bbb N}}

\def\Lb#1{\label{#1}}
\def\Ref#1{(\ref{#1})}

\def\[{\begin{eqnarray}}
\def\nn{\nonumber}
\def\non{\nonumber \\ }
\def\]{\end{eqnarray}}
\def\hh#1{\hspace*{#1}}

\def\een{\end{enumerate}}
\def\ben{\begin{enumerate}}

\def\FOR#1{\quad\mbox{for #1}}

\def\a{\alpha}

\def\d{\delta}
\def\e{\epsilon}
\def\f{\varphi}
\def\m{\mu}

\def\p{\psi}

\def\ba{\mbox{\boldmath$\alpha$}}
\def\bd{\mbox{\boldmath$\delta$}}
\def\be{\mbox{\boldmath$\eta$}}
\def\bL{\mbox{\boldmath$\Lambda$}}
\def\br{\mbox{\boldmath$\rho$}}

\def\bx{\mbox{\boldmath$\xi$}}
\def\bz{\mbox{\boldmath$\zeta$}}
\def\Dl{\Delta}
\def\L{\Lambda}

\def\cl{{\ell}}

\def\cV{{\cal V}}

\def\gg{{\got g}}
\def\gh{{\got h}}
\def\gn{{\got n}}
\def\su{{\got su}}

\def\sL{{\got L}}
\def\ew{{\got w}}
\def\eC{{\cal C}}

\def\eL{{\cal L}}
\def\eW{{\got W}}
\def\va{{\bf a}}
\def\val{{\ba}}
\def\vb{{\bf b}}
\def\vd{{\bd}}

\def\vk{{\bf k}}
\def\vl{{\bf l}}
\def\vL{{\bL}}
\def\vo{{\bf 0}}

\def\vr{{\bf r}}
\def\vri#1{{{\bf r}_{#1}}}

\def\vs{{\bf s}}
\def\vt{{\bf t}}

\def\vx{{\bf x}}

\def\vro{{\br}}
\def\vxi{{\bx}}
\def\8{{\bx_8}}

\def\vet{{\be}}
\def\vze{{\bz}}
\newfont\sgb{cmmib7} 

\def\Null{{\hbox{\sgb \symbol{"0E}}}}
\def\smxi{{\hbox{\sgb \symbol{"18}}}}

\def\smvL{{\hbox{\sgb \symbol{"03}}}}

\def\mult{{\rm mult}}
\def\ad#1#2{({\rm ad}\,#1)^{#2}}

\def\frc#1#2{{\textstyle \frac{#1}{#2}}}
\def\2{\frc12}

\def\9{{E_9}}
\def\0{E_{10}}

\def\|{\,|\,}
\def\.{\cdot}
\def\X{\!\cdot\!}

\def\res{{\rm Res}}

\def\1{{\bf 1}}

\def\Vp{\cV(\p,z)}

\def\ket{\rangle}
\def\ketr{|\vr\rangle}
\def\kets{|\vs\rangle}
\def\keta{|\va\rangle}

\def\amm{\a^\m_m}

\def\nod{\mbox{\bf:}}

\def\Ord{\phantom{.}_\times^\times}
\def\II{I\hspace{-.2em}I_{9,1}}
\def\III{I\hspace{-.2em}I_{25,1}}
\def\ggA{\gg(A)}

\def\ggL{\gg_\L}
\def\ggI{\gg_{\II}}

\begin{document}

\arraycolsep3pt

\thispagestyle{empty}
\renewcommand{\thefootnote}{\fnsymbol{footnote}}
\begin{flushright} hep-th/9411188 \end{flushright}
\vspace*{2cm}
\begin{center}
{\LARGE \sc $\0$ for beginners\footnote[1]{Invited
           talk presented by H. Nicolai at the Feza G\"ursey
        Memorial Conference, Istanbul, June 1994.}}\\
 \vspace*{1cm}
       {\sl Reinhold W. Gebert and Hermann Nicolai}\\
 \vspace*{6mm}
     IInd Institute for Theoretical Physics, University of Hamburg\\
     Luruper Chaussee 149, D-22761 Hamburg, Germany\\
 \vspace*{6mm}
\end{center}
\vspace*{1cm}

In this contribution, we summarize a recent attempt to understand
hyperbolic Kac Moody algebras in terms of the string vertex
operator construction \cite{GN} (which readers are also advised
to consult for a comprehensive list of references). As is well known,
Kac Moody algebras (see e.g.\ \cite{Kac90}, \cite{GodOli86})
fall into one of three classes corresponding to whether the associated
Cartan matrices are positive definite, positive semi-definite and
indefinite. Of these, the first two are well understood, leading to
finite and affine Lie algebras (the latter being equivalent to
current algebras in two space-time dimensions). Virtually nothing
is known, however, about the last class of Kac Moody algebras, based
on indefinite Cartan matrices. Nonetheless these have been
repeatedly suggested as natural candidates for the still elusive
fundamental symmetry of string theory (see e.g.\ \cite{Moor93},
\cite{West94} for recent proposals in this direction).
Being vastly larger than affine Kac Moody algebras,
they are certainly ``sufficiently big'' for this purpose,
but an even more compelling argument supporting such speculations
is the intimate link that exists between Kac Moody algebras
and the vertex operator construction of string theory which
has been known for a long time. More specifically,
it has been established that the elements making up a Kac Moody
algebra of finite or affine type can be explicitly realized in terms
of tachyon and photon emission vertex operators of a compactified
open bosonic string \cite{FreKac80}, \cite{GodOli85}. On the basis of
these results, it has been conjectured that Kac Moody algebras of
indefinite type might not only furnish new symmetries
of string theory, but might themselves be understood in terms
of string vertex operators associated with the higher excited
(massive) states of a compactified bosonic string \cite{GodOli85},
\cite{Fren85}.

Of what nature are these new symmetries then? In \cite{Gros88}, it was
argued that in the ultrahigh-energy
limit of string theory, where the Planck mass goes to zero, an
infinite number of linear relations exists between the scattering
amplitudes of different string states that are valid order by order
in perturbation theory. This suggests an enormous symmetry which is
restored at high energies. It may seem unreasonable to study
such a queer limit but, in fact, it is a very conservative approach.
For example, if we had not known the spontaneously broken
symmetry of the electroweak interactions, we could have in principle
discovered it at high energies where all gauge particles become
massless again. In agreement with this analogy it is indeed a reasonable
hope to find hints of the unbroken string gauge algebra by studying
the relations between high energy scattering amplitudes. But since
the latter, according to the above result, are essentially unique for
a given number of scattered physical states, it is tempting to
regard the Lie algebra of physical states itself as part of some
universal gauge algebra. Note that we obtain, by construction,
different Lie algebras of physical states when we consider
inequivalent string backgrounds. Moreover, due to the presence of
infinitely many massive physical states, each Lie algebra would
have to be spontaneously broken almost completely. If we take this
picture for granted then our task will be to make a
clever choice for the specific string background in order to find a
Lie algebra of physical states as large as possible. `Clever' here
apparently means `as symmetric as possible', and one is therefore
naturally led to Minkowskian torus compactifications where {\it all}
spacetime dimensions are chosen to be periodic (hence ``finite in all
directions'' \cite{Moor93}). More specifically, for the 26-dimensional
bosonic string there is a unique choice of maximal symmetry, namely
the even selfdual Lorentzian lattice $\III$ which indeed provides a
``large'' algebra --- the infinite rank fake monster Lie algebra
introduced by Borcherds \cite{Borc90}. To gain insight into the
mathematical structure of these symmetry algebras it is
instructive to study toy models such as the 10-dimensional bosonic
string compactified with momentum lattice $\II$.

The above, to some extent heuristic arguments were recently put on
more solid ground in \cite{West94}. In this paper, it was established
that the fake monster Lie algebra {\it is} a symmetry of
string theory in the sense that every physical state leads to a
symmetry of the string scattering amplitudes. In view of this
result one could now pose the question to which extent the vertices
are already fixed by stipulating the fake monster Lie algebra
as symmetry algebra. The degree of uniqueness would then give us
a clue of how small the algebra is in comparison with the universal
string gauge algebra. Certainly, they cannot be the same. For on the
one hand it is clear that the string vertices describe the string
field theory, on the other hand we know (see \cite{Moor93}) that the
fake monster Lie algebra does not contain all Lie algebras arising
from other string backgrounds. It is worth mentioning that the
calculations were carried out in the so-called group theoretical
approach to string theory which seems to be a powerful formalism to
analyze the issue of string symmetries.

Apart from possible relations to string theory, hyperbolic Kac Moody
algebras might appear in the dimensional reduction of
(extended) supergravity theories to one dimension \cite{Julia}.
Some evidence for this conjecture was presented in \cite{Nic}, where
it was argued that the Matzner Misner group arising in the reduction
of Einstein's theory to two dimensions can be generalized to a
``Matzner Misner ${\rm SL} (3, \R )$'' group providing precisely
the two nodes needed to extend the Dynkin diagram to a hyperbolic
one. The null Killing reduction required for this investigation
has been recently studied in \cite{JN}. We also mention that these
hidden symmetries may be related to S,T and U duality symmetries
arising in string theories (see \cite{Hull94} for recent progress
in this direction).

\medskip \par
Let us begin by reviewing how one constructs a Kac Moody algebra $\ggA$
from a given Cartan matrix $A = (a_{ij})$, where the indices $i,j$
are assumed to take $d$ values (so $d$ is the rank of this algebra),
and where the matrix $A$ may also be indefinite.
The basic building blocks are the Chevalley generators
$e_i, f_i, h_i$, which are subject to the following relations
\[ [h_i,e_j]&=&a_{ij}e_j,\quad [h_i,f_j]=-a_{ij}f_j, \\  {}
   [e_i,f_j]&=&\d_{ij}h_i. \]
The elements $h_i$ then automatically obey $[ h_i , h_j ] =0$
and constitute a basis of the Cartan subalgebra of
$\ggA$. The free Lie algebra associated with $A$
is obtained by forming multiple commutators of the elements
$\{e_i,f_i,h_i\,|\,i\}$ in all possible ways taking into account
the above relations. To obtain the Kac Moody algebra
$\ggA$ itself we must still divide out by the Serre relations
\[ \ad{e_i}{1-a_{ij}}e_j=0, \quad
   \ad{f_i}{1-a_{ij}}f_j=0. \Lb{Serre} \]
It is a standard result \cite{Kac90}
that this algebra can be written as a direct sum
\[ \ggA = \gn_+ \oplus \gh \oplus \gn_-  .         \]
The subalgebras $\gn_-$ and $\gn_+$ are defined to consist
of all linear combinations of multiple commutators of
the form $[f_{i_1},[f_{i_2},\ldots[f_{i_{n-1}},f_{i_n}]\ldots]]$ and
$[e_{i_1},[e_{i_2},\ldots[e_{i_{n-1}},e_{i_n}]\linebreak[0]
 \ldots]]$, respectively, modulo the multilinear Serre relations
\Ref{Serre}. Since the subalgebras $\gn_+$ and $\gn_-$ are conjugate
to each other, it is in practice sufficient to analyze either of them.
In this way one gets for positive definite or positive semi-definite
$A$ just the finite or affine Kac Moody algebras,
respectively (for the affine algebras, one still has to add an outer
derivation to $\gh$ due to the degeneracy of the Cartan matrix).
When $A$ is indefinite, on the other hand, the number of multiple
commutators ``explodes'', and no manageable way to handle
them is known that would be analogous to the
realization of affine Kac Moody algebras in terms of current algebras.
More specifically, the characteristic feature of
indefinite Kac Moody algebras is
the exponential growth in the number of Lie algebra elements
associated with a root $\vL$ as ${\vL}^2 \rightarrow - \infty$.
Thus for a given number of Chevalley generators
$e_{i_1},...,e_{i_n}$, the number of inequivalent ways of arranging
them into non-vanishing multiple commutators increases exponentially
with $-{\vL}^2$, where $\vL = \vr_{i_1} +\ldots+ \vr_{i_n}$
and $\vr_i$ are the simple roots associated with $e_i$.
This problem does not occur for either finite or affine Kac Moody
algebras, for which ${\vL}^2 =2$ or 0 are the only possibilities.
The problem here is not so much the enormous number of commutators,
but rather the fact that all those multiple commutators which contain
the Serre relation somewhere inside, or can be brought to
such a form by use of the Jacobi identities, must be identified
and discarded. Even the more modest
question as to how many elements there are for a given root
has defied all attempts at a general solution so far.
For a limited number of cases, one knows explicit
multiplicity formulas counting the dimensions of the root spaces,
but the complete root multiplicities are not known
for a single Kac Moody algebra of indefinite type
(root multiplicities can be determined in principle from the
Peterson recursion formula \cite{KMPS90}, but this formula quickly
becomes too unwieldy for practical use).

We are here mainly interested in {\it hyperbolic}
Kac Moody algebras which are based on indefinite $A$, but obey the
additional requirement that the removal of any point from the
Dynkin diagram leaves a Kac Moody algebra which is either of affine
or finite type. One can then show that the maximal rank is ten and
that the associated root lattice
must be Minkowskian, i.e.\ with metric signature $(+\ldots+|-)$. There
are altogether three hyperbolic algebras of maximal rank. Of these,
$\0$ is not only the most interesting, containing $E_8$ and its affine
extension $\9$ as subalgebras, but also distinguished by the
fact that it has only one affine subalgebra that can be obtained
by removing a point from the $\0$ Dynkin diagram, while the other
two rank 10 algebras contain at least two regular affine subalgebras.
Furthermore, the root lattice $Q(\0 )$
coincides with the (unique) 10-dimensional even unimodular Lorentzian
lattice $\II$ \cite{Con83}, whereas the root lattices of the
other two maximal rank hyperbolic algebras are not self-dual.
For the further discussion
it is useful to introduce the notion of {\it level}. Denoting the
``over-extended root'', which turns an affine into a hyperbolic
Kac Moody algebra, by $\vr_{-1}$ (e.g. for $\0$, this is the left-most
point in the Dynkin diagram, see below), one
defines the level $\cl \in \Z$ of a root such that positive
$\cl$ counts the number of $e_{-1}$ generators in
$[e_{i_1},[e_{i_2},\ldots[e_{i_{n-1}},e_{i_n}]\ldots]]$ (similarly, if
$\cl$ is negative, $-\cl$ counts the number of $f_{-1}$ generators
in $[f_{i_1},[f_{i_2},\ldots[f_{i_{n-1}},f_{i_n}]\ldots]]$). In terms
of the corresponding root $\vL = \vr_{i_1} +\ldots+ \vr_{i_n}$,
$\cl$ is defined by
\[   \cl := - \vL \X \vd, \]
where $\vd$ is the null vector of the affine subalgebra
obtained by deleting the over-extended node from the Dynkin diagram
(in principle one could also use the null vector of other regular
affine subalgebras to define a level which would be different from
the above; however, then not all of the results to be stated below
remain valid, e.g. the level-one states would no longer form
the basic representation of this affine subalgebra).
The level derives its importance from the fact that it grades the
hyperbolic Kac Moody algebra with respect to the
affine subalgebra \cite{FeiFre83}. Consequently, the
subspaces belonging to a fixed level can be decomposed into
irreducible representations of the affine subalgebra, the level
being equal to the eigenvalue of the central term on this
representation (the full hyperbolic algebra contains representations
of {\it all} integer levels). Multiplicities are known for levels
$\ell =0$ (corresponding to the affine subalgebra) and $\ell =1$
(corresponding to the so-called basic representation). More precisely,
we have \cite{Kac90}
\[  \mult(\vL) = p_{d-2} (1- \2 {\vL}^2 ),  \Lb{trans} \]
where $d$ is the rank of the algebra and the function $p_{d-2}(n)$
counts the partitions of $n\in\N$ into ``parts'' of $d-2$ ``colours''.
For $\cl =2$, one knows general multiplicity formulas
in some cases, and in particular for $\0$.
Beyond $\cl = 2$, no general formula seems to be known.
\medskip \par
Now it is well known that, at least
in principle, the string vertex operator construction can provide
a more concrete realization of an indefinite Kac Moody
algebra. To exploit it one interprets the root lattice as the
momentum lattice of a fully compactified string, and tries to
understand the multiple commutators in terms of string vertex
operators associated with the excited string states. The
real roots then correspond to spacelike (tachyon) momenta and the
imaginary roots to either lightlike or timelike momenta.
For the simple roots $\vr_i$, all of which
obey ${\vr_i}^2 =2$ (and hence are real), we have
the following correspondence
between Chevalley generators and tachyon and photon states:
\[ e_i&\mapsto& | \vr_i \ket , \\
   f_i&\mapsto& -| -\vr_i \ket, \\
   h_i&\mapsto&\vr_i(-1) | \vo \ket. \]
Here we use the shorthand notation $\vr_i(-1) \equiv \vr_i \cdot
{\val}_{-1}$ (where $\a_{-1}^\mu$
is just the lowest string oscillator);
furthermore, we define the states in such a way that
cocycle factors have been absorbed and do not appear explicitly.
The commutator between any two physical states $\f$ and $\p$
is quite generally defined by the formula (cf.\ \cite{Borc86})
\[    [ \p , \f ] := \res_z \left( \Vp \f \right),      \]
where $\Vp$ is the string vertex operator associated with the
state $\p$ (the residue formula here is completely equivalent
to the contour integrals employed down in \cite{GodOli85}).
An important consequence of this formula is that the physical
string states always form a Lie algebra (not to be confused with
the Kac Moody algebra, see remarks below).

Apart from yielding a concrete realization of an abstract algebraic
structure, the string vertex operators construction has the
advantage that the Serre relations \Ref{Serre} are built in
from the outset: in this context they simply state that the string
has no excited states ``below'' the tachyons.
For finite and affine Kac Moody algebras, no other states beside
tachyons and photons occur, whereas for indefinite $A$, excited
string states of arbitrarily high levels must be taken into account.
These will appear with certain polarizations, whose number increases
rapidly with the mass level of the string state. Thus, in more
physical parlance, the multiplicity of a root is equal to the
number of linearly independent polarizations of the corresponding
string state that can be reached by multiple commutators.
This ``easy'' realization of the Kac Moody algebra might suggest that
the problem of classifying the Lie algebra elements is essentially
solved by the string vertex operator construction.
Unfortunately, this is by no means the case because the problem
of accounting for the Serre relations is now replaced by another one:
not all physical states can be obtained in terms of
multiple commutators. Denoting the Lie algebra of {\it all} physical
states by $\ggL$, we rather have the {\it proper}
inclusion
\[ \ggA\subset\ggL .\Lb{inclus} \]
In other words, the Lie algebra of physical states is
well understood in physical terms, but the actual Kac Moody algebra
$\ggA$ is only a subalgebra thereof, and all the complications
now reside in the way in which $\ggA$ is embedded in the bigger, but
simpler algebra $\ggL$. In particular, there are ``missing
states'', i.e.\ physical states that cannot be represented as multiple
commutators of the Chevalley generators. A possible explanation
for this phenomenon is the following. For continuous momenta it
is well known that one can generate any physical state by multiple
scattering of tachyons (multiple scattering is the equivalent of
taking multiple commutators), so there can be no ``missing states''.
This is no longer the case for the compactified string: when the
tachyon momenta are on the lattice, certain ``intermediate states''
are no longer allowed, and therefore not all physical states are
accessible in this way any more. The construction given in \cite{GN}
is an attempt to make this intuitive argument more precise. As a
further consequence of \Ref{inclus}, the multiplicities are {\it not}
given by the numbers of excited states of the string (which are
of course well known), but only bounded above by them:
\[  {\rm mult} \vL \leq p_{d-1} ( 1 - \frc12 {\vL}^2 )
               - p_{d-1} ( - \frc12 {\vL}^2 ) . \Lb{bound} \]
Only for Euclidean lattices the two Lie algebras coincide, and
we have equality in \Ref{inclus}.

To summarize: the root system of the Kac Moody algebra $\ggA$ is well
understood though its root multiplicities are not completely
known for a single example; whereas the Lie algebra of physical
string states $\ggL$ has a much simpler structure (although it will
not be easy to define a root system associated with it).
Thus a complete understanding of \Ref{inclus} requires a
``mechanism'' which tells us  how $\ggA$ has to be filled up with
physical states to reach the complete Lie algebra of physical states.

We mention that recently Borcherds \cite{Borc88} has introduced a new
type of generalized Kac Moody algebras by admitting ``imaginary simple''
roots (i.e.\ obeying ${\vr_i}^2 \leq 0$). In the present
interpretion this means that one adds ``by hand'' those physical
states not obtainable as multiple commutators; the corresponding
momenta will then be counted as new simple roots, whose multiplicity
is simply given by the number of associated independent (missing)
polarizations. However, this program has so far only been carried
to completion for the 26-dimensional bosonic string, where special
properties such as the no-ghost theorem play a crucial role.
In this example, all missing states are under control (though not
explicitly known): one has to adjoin a certain (infinite) set of
photonic states as new Lie algebra generators to the ordinary
Kac Moody generators in order to get a complete set of generators
for the Lie algebra of physical states. The resulting algebra
constitutes an example of a generalized Kac Moody algebra and
has been dubbed fake monster Lie algebra (for historical reasons).
The imaginary simple roots corresponding to the extra generators
are just the positive integer multiples of the (lightlike) Weyl
vector for the lattice $\III$, and their multiplicities are equal to
the number of photon states (i.e.\ $=24$). On the other hand, for
algebras such as $\0$, not much is gained by this change of
perspective, because supplying the missing generators ``by hand''
presupposes knowledge of what the missing Lie algebra elements are
(not to mention the potential arbitrariness as to the number of ways
in which this can be consistently done). So the problem identifying
the elements belonging to $\ggL$ and not to $\ggA$ in \Ref{inclus}
remains.
\medskip \par
In \cite{GN} it is proposed to understand Kac Moody algebras
of hyperbolic type, and in particular the maximally extended
hyperbolic algebra $\0$, from a more ``physical'' (i.e.\ pedestrian)
point of view by focussing on the ``missing states'' rather than on
the Serre relations. For this purpose, we make use of a lattice
version of the DDF construction, which provides the most direct and
explicit solution of the physical state constraints in string theory
\cite{DeDiFu72}. The physical states, which by definition are
annihilated by the Virasoro constraints, are simply obtained in this
scheme by acting on a tachyonic groundstate with the DDF operators,
which commute with all Virasoro generators and form a spectrum
generating algebra. Our key observation is that for Kac Moody
algebras of ``subcritical'' rank (i.e.\ $d<26$), there appear {\it
longitudinal} string states and vertex operators beyond level one,
whose significance in this context has so far not been recognized to
the best of our knowledge. The appearance of longitudinal states is
already obvious from the known multiplicity formulas for level $\ell
=2$: for sufficiently large (negative) ${\vL}^2$, one can check
that there are roots $\vL$ such that
$\mult(\vL) > p_{d-2} (1-\2 {\vL}^2)$ (however, there may also exist
higher level roots whose multiplicity is {\it less} than the number of
transversal states). This clearly implies that whereas
transversal states are sufficient to characterize the elements
of an affine Kac Moody algebra (see below for further explanations
of this point), they are no longer sufficient for indefinite
Kac Moody algebras.

The necessary adaptation of the DDF construction
involves a discretization of the string vertex operator
formalism. As is well known \cite{GodOli85}, the allowed momenta
of the string excitations must be elements of the weight lattice
of the corresponding (affine or indefinite) Kac Moody algebra.
For the definition of DDF operators
one must choose a special Lorentz frame,
in terms of which one can distinguish transversal and longitudinal
DDF operators. On the lattice it is no longer possible to find a
frame such that the relevant DDF vectors (see below for details)
are still on the lattice, and we
therefore are forced  to make use of a rational extension of
the (self-dual) root lattice as an auxiliary device. This is a curious
feature of our construction, not encountered in previous studies.
Despite the fact that our DDF vectors are not on the lattice, we
employ them in our analysis
because we can still use the associated (transversal and
longitudinal) DDF operators to construct a complete basis for any
root space of the Lie algebra of physical states $\ggI$, of which
the corresponding root space of the Kac Moody algebra $\ggA$ is then
a proper subspace. As it turns out, longitudinal
states are absent only for
levels  zero and one; this accounts for the
comparative simplicity of the corresponding multiplicity formulas.

A central role in the DDF construction is played by
the tachyon momentum $\va$ (i.e.\ $\va^2=2$) of the ground state
$|\va \ket$, on which the physical states are built by means of
DDF operators, and a null vector $\vk$, subject to the condition
$\vk \X \va = 1$. For continuous momenta $\va$, we can always
find suitable $\vk = \vk (\va )$; moreover, we can rotate
these vectors into a convenient frame by means of a Lorentz
transformation \cite{Scher75}. On the lattice, however, the
full Lorentz invariance is broken to a discrete subgroup (containing
the Weyl group generated by the fundamental Weyl reflections), and
for generic roots $\vL$, the associated DDF vectors $\va$ and
$\vk$ will {\it not} be elements of the root lattice in general.
We can, however, still rotate the vectors $\va - n\vk$ into the
{\it fundamental Weyl chamber} for $n$ sufficiently large.
The lightlike momentum $\vk$ is then proportional to the
null-root $\vd$ characterizing the affine subalgebra; the latter
is always in the fundamental Weyl chamber. Now we invoke the
(obvious) fact that any root $\vL$ in the fundamental
Weyl chamber can be represented in the form
\[ \vL = \cl \vri{-1} + M \vd + \vb , \]
where $\cl$ is the level of $\vL$ and
$\vb$ an element of the $E_8$-root lattice $Q(E_8)$
($\vb$ need not be positive by itself as only $M \vd + \vb$
must be positive). We then define the {\bf DDF decomposition}
of $\vL$ by
\[ \vL = \va - n \vk(\va), \Lb{DDFdec1} \]
where
\[ \vk (\va ) := - \frc1{\cl} \vd  \Lb{DDFdec2} \]
and
\[ n = 1 - \2 \vL^2 = 1 + (M-\cl ) \cl - \2 \vb^2. \]
By construction, $\va$ obeys $\va^2 =2$ and is therefore associated
with a tachyon state, and $n$ is the number of steps required to
reach the root $\vL$ by starting from $\va$ and decreasing the
momentum by $\vk$ at each step ($n$ is non-negative because
${\vL}^2 \leq 2$; note also that $\vk$ is always a {\it negative} root,
so $\vL$ is positive for all $n$). Obviously, for
$\cl >1$, neither $\vk$ nor $\va$ belong to the lattice in general.
As a consequence, the intermediate DDF states associated
with momenta $\va -m \vk$ not on the lattice will not correspond
to elements of the algebra,
but they are nonetheless indispensable for the
construction of a complete basis for any given root space.
On the other hand, states
associated with the root $\vL$ do belong to the algebra of
physical states, and the DDF decomposition enables
us to write down all possible polarization states associated with the
root $\vL$ in terms of transversal and longitudinal DDF
states; the totality of these states constitutes the complete set of
elements in the root space $\ggI^{(\smvL)}$, whose dimension equals
$p_{d-1} (1- \frc12 {\vL}^2 ) - p_{d-1} (- \frc12 {\vL}^2 )$.
Explicitly, given a tachyon momentum $\va$, the physical states are
\[ A^{i_1}_{-m_1}(\va) \ldots A^{i_M}_{-m_M}(\va)
   \sL_{-n_1}(\va)\ldots \sL_{-n_N}(\va)\keta  , \Lb{DDFbasis} \]
explicitly indicating the dependence of the DDF operators
and their polarizations on the tachyon momentum $\va$ and the
associated lightlike vector $\vk (\va) = -\frc1{\ell} \vd$,
and assuming $n_i \geq 2$ to exclude null states.
The transversal DDF operators associated with
a tachyonic momentum $\va$ and its light-like momentum $\vk$
are defined by the well known formula
\[  A^i_{-m} = \res_z \big[ \cV ( \vxi_i (-1)
 | - m\vk \ket , z ) \big], \Lb{DDFdef}          \]
i.e.\ is given by the contour integral of the vertex operator
corresponding to the photon state $\vxi_i \X {\val}_{-1}
| \frc{m}{l} \vd \ket $.
The transversal DDF operators always form a $(d-2)$-fold Heisenberg
algebra. Note, however, that we have to deal with a whole plethora
of transversal Heisenberg algebras, namely one for each admissible
pair $(\va,\vk)$; this is in contrast to the single set of primitive
oscillators, $\{\amm|1\le\mu\le d, m\in\Z\}$, which makes up the
full Fock space when acting on the groundstates.
The longitudinal DDF operators are given by
more complicated (but also standard) expressions; they involve
a logarithmic contribution (cf.\ \cite{Brow72}).
Again, for each admissible pair $(\va,\vk)$, we end up with a
different set of operators.
An important technical point is that the longitudinal
vertex operators cannot be associated with definite states, as their
action cannot be defined on all of Fock space, including the
vacuum state $|\vo\ket$. Put differently,
they do not correspond to summable operators in the sense of
\cite{FLM88}; in this respect, vertex algebras
encompassing longitudinal states transcend
the definition given in \cite{Borc86}, \cite{FLM88}.

We quickly summarize some pertinent results about $\0$.
It is defined from its Cartan matrix in terms of multiple
commutators of the Chevalley generators as described above.
The root lattice $Q(\0 )$ of $\0$ is the unique even self-dual
$\II$ in ten dimensions, which can be defined as the set of points
$\vx=(x_1,\ldots,x_{9}|x_0)$ in 10-dimensional
Minkowski space for which the $x_\m$'s are all in $\Z$ or
all in $\Z+\frac12$ and which have integer inner product with the
vector $\vl=(\frac12,\ldots,\frac12\,|\,\frac12)$, all norms and inner
products being evaluated in the Minkowskian metric
$\vx^2=x_1^2+\ldots+x_9^2-x_0^2$ (cf.\ \cite{Serr73}).
According to \cite{Con83}, a set of positive norm simple
roots for $\II$ is given by the ten vectors $\vri{-1}, \vri0, \vri1,
\ldots, \vri8$ in $\II$ for which $\vr_i^2=2$ and $\vr_i\X\vro=-1$
where the Weyl vector is $\vro=(0,1,2,\ldots,8|38)$ with
$\vro^2=-1240$.
The corresponding Coxeter-Dynkin diagram looks as follows
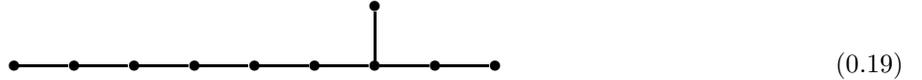
\[ \unitlength1mm
   \begin{picture}(66,10)
   \multiput(1,0)(8,0){9}{$\bullet$}
   \put(49,8){$\bullet$}
   \multiput(2.5,.9)(8,0){8}{\line(1,0){6.5}}
   \put(49.9,1.5){\line(0,1){6.5}}
   \end{picture}   \]
and is associated with the Cartan matrix:
$$ A\equiv (a_{ij})=(\vr_i\X\vr_j)=\left(
   \begin{array}{rrrrrrrrrr}
              2 &-1 & 0 & 0 & 0 & 0 & 0 & 0 & 0 & 0  \non
             -1 & 2 &-1 & 0 & 0 & 0 & 0 & 0 & 0 & 0  \non
              0 &-1 & 2 &-1 & 0 & 0 & 0 & 0 & 0 & 0  \non
              0 & 0 &-1 & 2 &-1 & 0 & 0 & 0 & 0 & 0  \non
              0 & 0 & 0 &-1 & 2 &-1 & 0 & 0 & 0 & 0  \non
              0 & 0 & 0 & 0 &-1 & 2 &-1 & 0 & 0 & 0  \non
              0 & 0 & 0 & 0 & 0 &-1 & 2 &-1 & 0 &-1  \non
              0 & 0 & 0 & 0 & 0 & 0 &-1 & 2 &-1 & 0  \non
              0 & 0 & 0 & 0 & 0 & 0 & 0 &-1 & 2 & 0  \non
              0 & 0 & 0 & 0 & 0 & 0 &-1 & 0 & 0 & 2  \nn
   \end{array} \right). $$

Let us first describe the $\9$ subalgebra in terms
of physical states.
The Cartan subalgebra of $\0$ (and also of $\9$) is spanned by
the states
\[ \vd(-1)|\vo\ket &=:&K, \Lb{E9-3} \\
   (\vr_{-1}+\vd)(-1)|\vo\ket &=:&d, \Lb{E9-4} \\
   \vxi_i(-1)|\vo\ket & &\FOR{}\ i=1,\ldots,8\ , \Lb{E9-5} \]
where $K$ represents the central element, $d$ is the derivation of $\9$,
and \(\{\vxi_i(-1)|\vo\ket \,|\, i=1,\ldots,8\}\) span the Cartan
subalgebra of $E_8$. This is the standard ``light-cone'' basis of
$\gh(\9)$ in the sense that $K$ and $d$ are lightlike.
The allowed (positive and negative) roots are all $\vr\in\II$ obeying
$\vr^2=2$ and $\vr\X\vd=0$ (hence having no $\vri{-1}$ component),
and $m\vd$ for $m \in \Z^\times$. These correspond to the tachyonic
states $|\vr\ket$ and the photonic states $\vxi_i(-1)|m\vd\ket$
(where $\vxi_i\X\vd= \vxi_i\X\vri{-1}= 0$) with multiplicities $1$ and
$8$, respectively. The following commutation relations are already
enough for a complete characterization of $\9$
\[ \big[\vet(-1)|\vo\ket ,\vze(-1)|\vo\ket \big]&=&0, \\[.5em]
   \big[\vet(-1)|\vo\ket ,\vxi_i(-1)|m\vd\ket \big]
       &=&m(\vet\X\vd)\vxi_i(-1)|m\vd\ket , \\[.5em]
   \big[\vet(-1)|\vo\ket ,\ketr \big]&=&(\vet\X\vr)\ketr , \\[.5em]
   \big[\vxi_i(-1)|m\vd\ket ,\vxi_j(-1)|n\vd\ket \big]
       &=&m\d_{m+n,0}(\vxi_i\X\vxi_j)\vd(-1)|\vo\ket , \\[.5em]
   \big[\vxi_i(-1)|m\vd\ket ,\ketr \big]
       &=&(\vxi_i\X\vr)|\vr+m\vd\ket , \\[.5em]
   \big[\ketr ,\kets \big]
       &=&\cases{0  & if $\vr\X\vs\ge0$, \cr
                 \e(\vr,\vs)|\vr+\vs\ket   & if $\vr\X\vs=-1$, \cr
                 -\vr(-1)|m\vd\ket   & if $\vr+\vs=m\vd$,} \]
for $\vet,\vze\in\gh(\9)$ and $\9$ roots $\vr,\vs$. In the last
formula $\e$ denotes the cocycle factor.
The bra and ket notation used here may appear a bit unusual (and is
not really necessary at this point), but will prove invaluable
as one goes to higher level elements of the hyperbolic algebra.
The multiplicities of the corresponding Lie algebra elements
can be read off directly, and are given by 1 and 8, respectively,
for the tachyonic and lightlike roots in accord with the
formula \Ref{trans} above.

The level-one elements exhibit already a slightly more involved
structure. Inspection of the inverse
Cartan matrix shows that the only such roots
in the fundamental Weyl chamber $\eC$ are of the form (for $k_{-1}
   \in \N$)
\[ \vL = \vr_{-1} + (2+ k_{-1}) \vd,  \]
corresponding to the DDF decomposition \Ref{DDFdec1} with
$\va = \vr_{-1}$, $\vk = - \vd$ and $n=2+ k_{-1}$. Since all these
vectors are elements of the lattice, we can straightforwardly apply
the DDF construction to obtain the physical states
\[ A_{-m_1}^{i_1}\cdots A_{-m_N}^{i_N}|\vr_{-1}\ket, \Lb{E10-lev1}  \]
where $m_1+\ldots+m_N=2+k_{-1}$ and with the polarization vectors
chosen such that $\vxi_i\X\vxi_j=\d_{ij}$ and $\vxi_i\X\vd=
\vxi_i\X\vri{-1}= 0$ for $i,j= 1,\ldots,8$. In terms of
multiple commutators, these states correspond to
\[   \big[\vxi_{i_1}(-1) | m_1 \vd \ket  ,
     \big[\ldots,
     \big[\vxi_{i_N}(-1) | m_N \vd \ket  ,
       | \vr_{-1} \ket  \big]\ldots\big]\big]\qquad\in\0^{(\smvL)}. \]
All relevant level-one states can now be obtained by acting with
the $\9$ Weyl group on these states and polarizations. Denoting the
rotated DDF operators by
$A^{\ew(i)}_{-m}\equiv A^{\ew(\smxi_i)}_{-m}$, we obtain the new states
\[ A_{-m_1}^{\ew(i_1)}\cdots A_{-m_N}^{\ew(i_N)}|\ew(\vr_{-1})\ket
                                          \Lb{basic}                \]
in this fashion.
The so-called basic representation is spanned by all elements
of the form \Ref{basic}; the highest weight vector of the representation
is easily seen to be $|\vri{-1}\ket$.
\medskip \par
At level two, a general multiplicity formula was derived in
\cite{KaMoWa88}; it reads
\[  \mult(\vL) = \xi (3- \2 {\vL}^2), \Lb{mult2}  \]
where
\[  \sum_{n \geq 0} \xi (n) q^n =
\frac{1}{\phi (q)^8}
\left[ 1- \frac{\phi (q^2)}{\phi (q^4)} \right], \]
and $\phi (q)$ is the Euler function. The special example we have
investigated in \cite{GN}, is the level-two root
$\vL = \vL_7$, dual to the simple root $\vri7$, which has $\vL_7^2= -4$
and is explicitly given by
\[ \vL_7=\left[\begin{array}{c@{\hh{.8em}}c@{\hh{.8em}}c@{\hh{.8em}}
                             c@{\hh{.8em}}c@{\hh{.8em}}c@{\hh{.8em}}
                             c@{\hh{.8em}}c@{\hh{.8em}}c}
                &   &   &   &    &    &  7 &   &   \\
               2& 4 & 6 & 8 & 10 & 12 & 14 & 9 & 4
               \end{array} \right]
        =(0,\ 0,\ 0,\ 0,\ 0,\ 0,\ 0,\ 0,\ 0\,|\,2), \]
where the first notation with square brackets refers to the simple
roots in the above Dynkin diagram. We can now invoke the results of
\cite{FeiFre83} which tell us that level $\ell$ states can be
obtained as $(\ell -1)$-fold commutators of level-one states, for
which we can use the representation \Ref{basic}. Our analysis reveals
that the following states form a complete basis of the root space
$E_{10}^{(\smvL_7)}$ (no summation convention!):
\[ A^i_{-2}A^j_{-1}\keta &&
   \FOR{$i,j$ arbitrary,} \non
   A^i_{-1}A^j_{-1}A^k_{-1}\keta &&
   \FOR{$i\neq j\neq k \neq i$,} \non
   \big\{A_{-3}^i - A^i_{-1}A^j_{-1}A^j_{-1} \big\}\keta &&
   \FOR{$i \neq j$,} \non
   \big\{5A_{-3}^i + A^i_{-1}A^i_{-1}A^i_{-1}\big\}\keta &&
   \FOR{$i$ arbitrary,}  \non
   \big\{A_{-3}^i - A^i_{-1} \sL_{-2}\big\}\keta &&
   \FOR{$i$ arbitrary.}  \Lb{E10-lev2}  \]
Altogether, we get $64 + 2\X56 + 2\X8 = 192 $
states in agreement with the formula \Ref{mult2} predicting
$\xi (3) = 192$ \cite{KaMoWa88}. Despite the fact that this number
coincides with the number of transversal states, our result
explicitly shows the appearance of longitudinal as well as the
disappearance of some transversal states. The above states form
irreducible representations of the little group;
in particular, the longitudinal DDF operator is inert under the
little Weyl group.
To appreciate the simplicity of this result readers need only
contemplate the problem of classifying the states
in terms of 75-fold multiple commutators of the Chevalley
generators for this example.

Having a complete description of the root space $E_{10}^{(\smvL_7)}$,
we can now in principle explore root spaces associated with
other level-two roots of the form $\vL = \vL_7 + n\vd$ (i.e.\ the
{\bf root string} associated with $\vL_7$) by commuting the states
\Ref{E10-lev2} with the $\9$ elements $\vxi_i(-1)|m\vd\ket$. From
\Ref{DDFdef} it is evident that all states obtained by
acting with a product $A^{i_1}_{-2m_1} \ldots A^{i_M}_{-2m_M}$
on any of the states \Ref{E10-lev2} belong to the
root space of $\vL = \vL_7 + (m_1 + \ldots m_M) \vd$ (note that
each operator $A^i_{-2m}(\va )$ shifts the momentum by $m\vd$!).
However, it is also clear that we cannot obtain all root space
elements in this way. For this, it is necessary to calculate
DDF commutators between appropriate elements of the form \Ref{basic}.
An alternative, more elucidating way might be to consider the action
of the Sugawara generators defined by
\[   \eL^{{\rm Sug.}}_m :=  \frac{1}{2(\ell+h^\vee)}
  \bigg\{\sum_{n\in\Z}\sum_{i=1}^8 \nod A^i_n A^i_{m-n} \nod
  +\sum_{\vs\in\Dl^{\rm real}(\9)}  \Ord{\rm ad}_{|\vs\ket}\,
     {\rm ad}_{|-\vs-m\Null\ket} \Ord \bigg\}   \]
on the states \Ref{E10-lev2}; here, $h^\vee=30$ is the dual Coxeter
number of $E_8$, the level is $\ell =2$, and the normal-ordering of
the operators ${\rm ad}_{|\vr\ket}$ is chosen as
\[ \Ord{\rm ad}_{|\vs+m\Null\ket}\,{\rm ad}_{|\vt+n\Null\ket}\Ord
                 :=\cases{
     {\rm ad}_{|\vs+m\Null\ket}\,{\rm ad}_{|\vt+n\Null\ket}
                          & if $m\ge n$, \cr
     {\rm ad}_{|\vt+n\Null\ket}\,{\rm ad}_{|\vs+m\Null\ket}
                          & if $m<n$, \cr} \]
for $E_8$ roots $\vs,\vt$ and $m,n\in\Z$. We get
\[ \eL^{\rm Sug.}_m \keta  = 0  \]
for $m\geq 1$. Furthermore, when evaluating $\eL^{\rm Sug.}_0$
on the ground state $\keta $, we find
$A_0^8\keta  = -2\keta $, and obtain
\[  \eL^{\rm Sug.}_0 \keta  = \frc{1}{16} \keta  , \]
showing that the state $\keta $ is a highest weight vector
of weight $h =\frc1{16}$ for the level-two Sugawara generators.
In view of the results of \cite{KaMoWa88}, we
therefore expect these states to belong to the irreducible
Virasoro module with $c=\frc12$ and $h=\frc1{16}$.
The problem that remains is to relate the Sugawara generators
to the longitudinal DDF operators. If this can be done, a completely
explicit description of {\it all} level-two root spaces is within
reach.
\medskip \par
Let us finally return to the issue of missing states in more detail.
Comparing \Ref{trans} with \Ref{bound}, it becomes obvious that
tachyonic and photonic physical states are necessarily transversal,
so that
\[ \ggI^{(\smvL)}\equiv\0^{(\smvL)}\FOR{}\vL^2\ge0  \]
(of course, for $\vL^2>2$, both spaces are empty). This means that
there are no missing states for $\vL^2\ge0$. But already for the
massive spin 2 states, we encounter one longitudinal physical state
that surely does not belong to the Kac Moody algebra $\0$. It is
clear that there is only one weight in $\eC$ of norm $-2$, namely
the fundamental weight $\vL_0=\vri{-1}+2\vd$. Since the latter is
a level-one element, which we know to occur in $\0$ just with
transversal polarizations, we infer that the longitudinal state
$\sL_{-2}|\vri{-1}\ket$ is not contained in the root space
$\0^{(\smvL_0)}$ and thus represents a missing state, so
\[ \ggI^{(\smvL_0)}
                \equiv\0^{(\smvL_0)}+\R\X\sL_{-2}|\vri{-1}\ket. \]
Acting with the full Weyl group on the missing state, we obtain
the associated orbit of missing states in $\0$. Our detailed analysis
of the root space for $\vL_7$ now enables us to discuss the case of
norm $-4$, for $\vL_7$ is the only weight in the fundamental Weyl
chamber with this property. From the multiplicity formula we learn
that there have to be $201-192=9$ missing states, and in view of our
DDF basis \Ref{E10-lev2} we write
\[ \ggI^{(\smvL_7)}\equiv\0^{(\smvL_7)}
                    +{\rm span}_\R\{\sL_{-3}|\va\ket;\
                                 A^i_{-3}|\va\ket,i=1,\ldots,8\}, \]
which can be also acted on with $\eW(\0)$ to find its analogue in
other chambers.

The above formulas naturally suggest two ways of how to proceed.
If we are primarily interested in $\0$, we can try to systematize
the way of splitting of $\ggI$ into $\0$ states and missing states.
In other words, we are seeking a mechanism which satisfactorily
answers the following question: How do the missing states decouple
from the $\0$ states? That this idea is not far-fetched shows the
example of the 26-dimensional bosonic string. There we separate the
longitudinal physical states from the transversal ones by introducing
a positive semidefintite contravariant bilinear form which renders
the former states to be null physical states. If one prefers the
modern cohomological treatment then the decoupling is furnished by the
nilpotent BRST operator and its cohomology. Thus we may rephrase
the above question as: {\it Is there a cohomology describing how
$\0$ is embedded in the Lie algebra of physical states, $\ggI$?}

The other point of view, as advocated by Borcherds, involves a
generalization of the framework of Kac Moody algebras. We know how
a large part of $\ggI$, namely the $\0$ part, can be formulated in
terms of generators and relations. The idea then is to extend this
approach to the whole Lie algebra. We would have to find an additional
set of Chevalley generators which, when adjoined to the generators
for $\0$, produce all physical states as multiple commutators.
For example, we certainly have to add $\sL_{-2}|\vri{-1}\ket$ as such
a new generator. This amounts to saying that $\vL_0$ constitutes
an imaginary simple root with multiplicity one. Kac Moody algebras
allowing for imaginary ($\equiv$ nonpositive norm) simple roots were
invented by Borcherds \cite{Borc88}. So far, the introduction of this
new generator seems to be very natural and appealing, but the second
step of the procedure is subtle and becomes cumbersome when repeatedly
done. In order to decide which missing states for the case of
$\vL_7$ have to be chosen as new generators, we need to take
into account the previous additional generator $\sL_{-2}|\vri{-1}\ket$.
Thus we ought to calculate the commutator $[|\vs\ket,\sL_{-2}|\vr\ket]$
(where $\vs+\vr+2\vd=\vL_7$) and express it in terms of the DDF basis
for $\ggI^{(\smvL_7)}$ to see which missing states now do appear. We
have not completed this calculation yet, for we are mainly interested
in $\0$ itself and hence focus on the first approach. Alternatively,
it is also possible to determine recursively the imaginary simple
roots by anticipating $\ggI$ as a Borcherds algebra and then plug its
well-known root multiplicities, $p_9(1-\2\vr^2)-p_9(-\2\vr^2)$,
into the Weyl--Kac--Borcherds denominator formula \cite{Borc88}.

So, what have we learnt from our analysis of the root space
$\0^{(\smvL_7)}$ and how may it be relevant for other hyperbolic
Kac Moody algebras? Our approach suggests that root spaces of $\0$
and other algebras of that type carry an additional structure related
to polarization; this differs from the conventional point of view
that a root space is essentially, up to its dimension, a black box.
The DDF framework, as developed here, provides adequate tools for
the analysis of the complicated structure of hyperbolic algebras.

In particular, we now have a deeper understanding why Frenkel's
conjecture \cite{Fren85} is wrong. Inspired by the example
of the 26-dimensional bosonic string and the results about the
canonical hyperbolic extension of $\su(2)$ \cite{FeiFre83}, he
conjectured that for every hyperbolic algebra $\gg$ of rank $d$ one
has, for any root $\vr$, $\dim\gg^{(\vr)}\le p_{d-2}(1-\2\vr^2)$
as an upper bound. This conjecture was disproved in \cite{KaMoWa88}
by establishing the level-two multiplicity formula for $\0$ as a
counterexample. We argue that the 26-dimensional bosonic
string represents a rather untypical example, because there the
longitudinal states span the radical of the contravariant bilinear
form which is divided out. Hence only transversal states survive and
we end up indeed with the exact multiplicity formula
$p_{24}(1-\2\vr^2)$. In the generic case, on the other hand, the
longitudinal states do appear as Lie algebra elements.
In terms of the DDF realization the following picture emerges. At
level-zero and level-one we naturally obtain all
transversal states giving the affine subalgebra and its basic
representation, respectively. By commuting transversal level-one
states, which is necessary for generating higher level elements,
we cannot escape from producing longitudinal states, too. Hence there
is no reason to expect a connection between higher level root
multiplicities and the formula $p_{d-2}(1-\2\vr^2)$, which just
counts the number of transversal states. Of course, we start off from
the transversal level-one states, but the more commutators we take
between them the more subtle the entanglement of longitudinal
and transversal states becomes.

For example, look at the canonical hyperbolic extension of $\su(2)$
whose level-two root multiplicities coincide with the number of
transversal states, $p_1(1-\2\vr^2)$, up to $\vr^2\le-36$ (see
\cite[Table $H_3$]{Kac90}) and then drop below this bound. We conjecture
that, when we perform the DDF construction for this example, we shall
see at level-two from the very beginning longitudinal states to appear
and transversal states to be missed, even though the multiplicity
superficially suggests the existence of transversal states alone.
For higher levels we predict an increasing mixing of longitudinal
and transversal states which manifests itself in an increasing
deviation of the multiplicities from the number of transversal states.
Thus the DDF analysis of a single level-two root space of $\0$
allows us to make some reasonable predictions for some features
occurring in other hyperbolic algebras of that type.

\end{document}